\documentclass[a4paper,11pt,sort&compress,number,5p]{elsarticle}
\usepackage[utf8]{inputenc}
\usepackage{lineno}
\usepackage{pifont}
\usepackage{natbib}
\usepackage{fleqn}
\usepackage{graphicx}
\usepackage{txfonts}
\usepackage{multicol}
\usepackage{multirow}
\usepackage{caption}
\usepackage{tabularx}
\usepackage{floatflt}
\usepackage{xcolor}
\usepackage{booktabs}
\usepackage{hyperref}
\usepackage{dcolumn}
\modulolinenumbers[5]
{}
\journal{Materials Today Communications}

\begin{document}
		
\begin{frontmatter}
\title{Effect of doping on SGS and weak half-metallic properties of inverse Heusler Alloys}
\author[1]{R.~Dhakal}
\ead{ramesh.dhakal91@gmail.com}
\author[1]{S.~Nepal}
\ead{sashinepal@gmail.com}
\author[1,2]{R. B. Ray}
\ead{ray\_rb@ymail.com}
\author[3]{R. Paudel}
\ead{gck223@gmail.com}
\author[1]{G. C. Kaphle\corref{cor1}}
\ead{ramesh.dhakal91@gmail.com}
\cortext[cor1]{Corresponding author}
\address[1]{Central Department of Physics, Tribhuvan University, Kathmandu, Nepal}
\address[2]{Department of Physics, Amrit Campus, T. U., Kathmandu, Nepal}
\address[3]{School of Materials Science and Engineering, Harbin Institute Of Technology, Harbin, 150001,China}
\begin{abstract}
Heusler alloys with Mn and Co have been found to exhibit interesting electronic and magnetic properties. Mn$_2$CoAl is well known SGS compound while Mn$_2$CoGa has weak half metallic character. By using plane wave pseudo-potential method, we studied the effect of Fe and Cr doping on half-metalicity and magnetism of these compounds. The doping destroys the SGS nature of  Mn$_2$CoAl  while the small-scale doping enhance the half-metallicity of Mn$_2$CoGa making it perfect half-metal. In case of Mn$_2$CoAl, the doping decrease the band gap while increase in band width is noticed for Mn$_2$CoGa. The half-metallicity is destroyed in both cases when the doping level is beyond certain degree. Moreover, we have also computed magnetic behavior of Mn$_2$CoZ alloys and we found that total magnetic moments of dopped samples have higher values than that of pristine compounds.            
\end{abstract}
\end{frontmatter}
\section{Introduction}
In last two decades, Heusler alloys have been studied extensively because of its diverse magnetic phenomena. Spin gapless \cite{cite1}  and half metallic properties\cite{2} have been reported and SGS is verified experimentally\cite{3} in Heusler family, the importance of which lies in its potential application in the realm of spintronics. The two spin channels show entirely contrasting behavior in half metallic and spin gapless materials which helps to manipulate the charge carriers and should enhance the performance of magnetoelectronic devices such as giant magnetoresistance spin-valves\cite{a2}, magnetic tunneling junctions\cite{a3}, spin-injecting\cite{a4} and spin transfer torque  devices\cite{a5}. {{Moreover, the research on Heusler alloys optical properties have also drawn significant attention though optoelectronic properties is widely studied in perovskite\cite{a9,a10,a11}}.}
\par {{Various theoretical approach to study the phase stability\cite{a6}, thermal, mechanical and structural properties\cite{a7}\cite{a8} of Heusler alloys has appeared in literature but} Heusler} alloys containing Mn and Co have attracted particular attention because of the high Curie temperature and peculiar behavior of magnetic moment of Manganese atom. The neighboring atoms dictate the magnetic moment alignment of Manganese atom\cite{2}\cite{4} and the exchange interaction between Mn and Co is found to be short-range effect\cite{5}. The first experimental study of Mn$_2$CoZ compound was made by Liu \textit{et al.}\cite{2}, where they predicted Mn$_2$CoAl as half metallic and Mn$_2$CoGa as weakly half-metallic. Later it was revealed that Mn$_2$CoAl possesses spin gapless property\cite{3} whereas weak half-metallic behavior of Mn$_2$CoGa was corroborated by further investigation\cite{6}. Mn$_2$CoZ compounds crystallize in inverse Heusler structure with space group F$\bar{4}3$m(216) \cite{c1,c2}, where the sequence of atom is Mn-Mn-Co-Z on the diagonal basis, the structure prototype of Hg$_2$CuTi alloy. In this symmetry, the Mn atoms occupy A(0,0,0) and B($\frac{1}{4}$,$\frac{1}{4}$,$\frac{1}{4}$) sites, whereas Co and Z atoms take C($\frac{1}{2}$,$\frac{1}{2}$,$\frac{1}{2}$) and D($\frac{3}{4}$,$\frac{3}{4}$,$\frac{3}{4}$) sites respectively in Wyckoff positions\cite{d1}.
\par The theoretical explanation of the effect of doping and disorder on the properties like SGS and half-metallicity of Heusler alloys has appeared in literature \cite{7}\cite{8} but little attention has been given to the weak-half metallic character and the effect of doping on this particular property. In our work, we report the effect of doping on SGS and weak half metallicity of inverse Heusler alloy. For this purpose we chose the conventional cell of Mn$_2$CoZ (Z=Al, Ga) compounds, where the partial doping is done  by Fe and Cr at two different levels. The site occupation of the dopant atoms is determined by the general empirical rule as described in literature\cite{4} and the unusual site occupation of chromium has also been considered\cite{9}.
\par The general rule is if the dopant atoms have more valance electrons than that of Manganese atoms, then it will occupy A and C sites whereas if the valance electrons of dopant is less than that of Mn atoms, B and D sites will be preferred. But the Cr atoms as a dopant do not obey this rule and prefer to occupy A and C sites. So in our case, both Fe and Cr atoms will occupy the A sites whereas the displaced Mn(A) atoms at A sites will replace the Al(D) or Ga(D) atoms at D sites. During
this modeling, it is assumed that the lattice constant of a crystal has not changed due to doping. Also, it is assumed that structural phase transition has not taken place so that the crystal remains in a cubic structure.
\section{Computational Details} 

The electronic and magnetic properties were calculated using plane-wave pseudopotential method within DFT framework implemented using 
Quantum ESPRESSO \\package\cite{11,12}. In our work, for Manganese, Aluminium and Iron atoms Projector-Augmented Wave(PAW) set was used and for Cobalt and Chromium atoms ultra-soft pseudo-potential was exploited. Exchange-correlation functional was approximated using Ernzerhof generalized gradient approximation (PBEsol GGA)\cite{13}\cite{b1,b2}. Cutoff energy for plane-wave was set to 160 Ry for Mn$_2$CoGa and 175 Ry for Mn$_2$CoAl. For k-point sampling in electronic structure calculation, we used Monkhorst-pack grid of size 8x8x8 for primitive cell and 6x6x6 for conventional cell. Linear tetrahedral integration method is used to calculate density of states\cite{14}. Lattice parameter was optimized using total energy minimization method. For doping, we used conventional structure with 16 atoms/cell. Coordinate relaxation was performed until total energy change between two consecutive scf step is less than 1x10$^{-5}$ Ry. Force convergence threshold was set to 1x10$^{-3}$ Ry/a.u. 

\section{Results and Discussion}
{The structural stability of compounds under study is confirmed by the calculation of formation and cohesive energy. The formation and cohesive energy for \\${Mn_2CoM_xZ_{1-x}}$[M=Cr,Fe \& Z=Al,Ga \& x=0,0.25,0.5] can be written as
\begin{equation}
\label{eq1}
E_{Form}^{Mn_2CoM_xZ_{1-x}}=E_{tot}-(2E_{Mn}^B+E_{Co}^B+E_M^B+E_Z^B)
\end{equation}
\begin{equation}
\label{eq2}
E_{Coh}^{Mn_2CoM_xZ_{1-x}}=E_{tot}-(2E_{Mn}+E_{Co}+E_M+E_Z)
\end{equation}
Here, $E_{Form}^{Mn_2CoM_xZ_{1-x}}$ and $E_{Coh}^{Mn_2CoM_xZ_{1-x}}$ represent the formation and cohesive energy of the compounds respectively whereas $E_{tot}$ is the total energy per formula unit. In equation \ref{eq1}, $E_{Mn}^B$, $E_{Co}^B$, $E_M^B$ and $E_Z^B$ indicate total energy per atom for bulk whereas in equation \ref{eq2} $E_{Mn}$, $E_{Co}$, $E_M$ and $E_Z$ represents the total energy of corresponding isolated atom. From table \ref{tab:tableis}, we can see that cohesive energy of all the compounds, pristine as well as doped sample, are negative which confirms the structural stability of our compounds.     }
\par
\begin{figure}[h] 
	\centering
	\label{Fig:1}
	\includegraphics[width=1\linewidth, height=0.37\textheight]{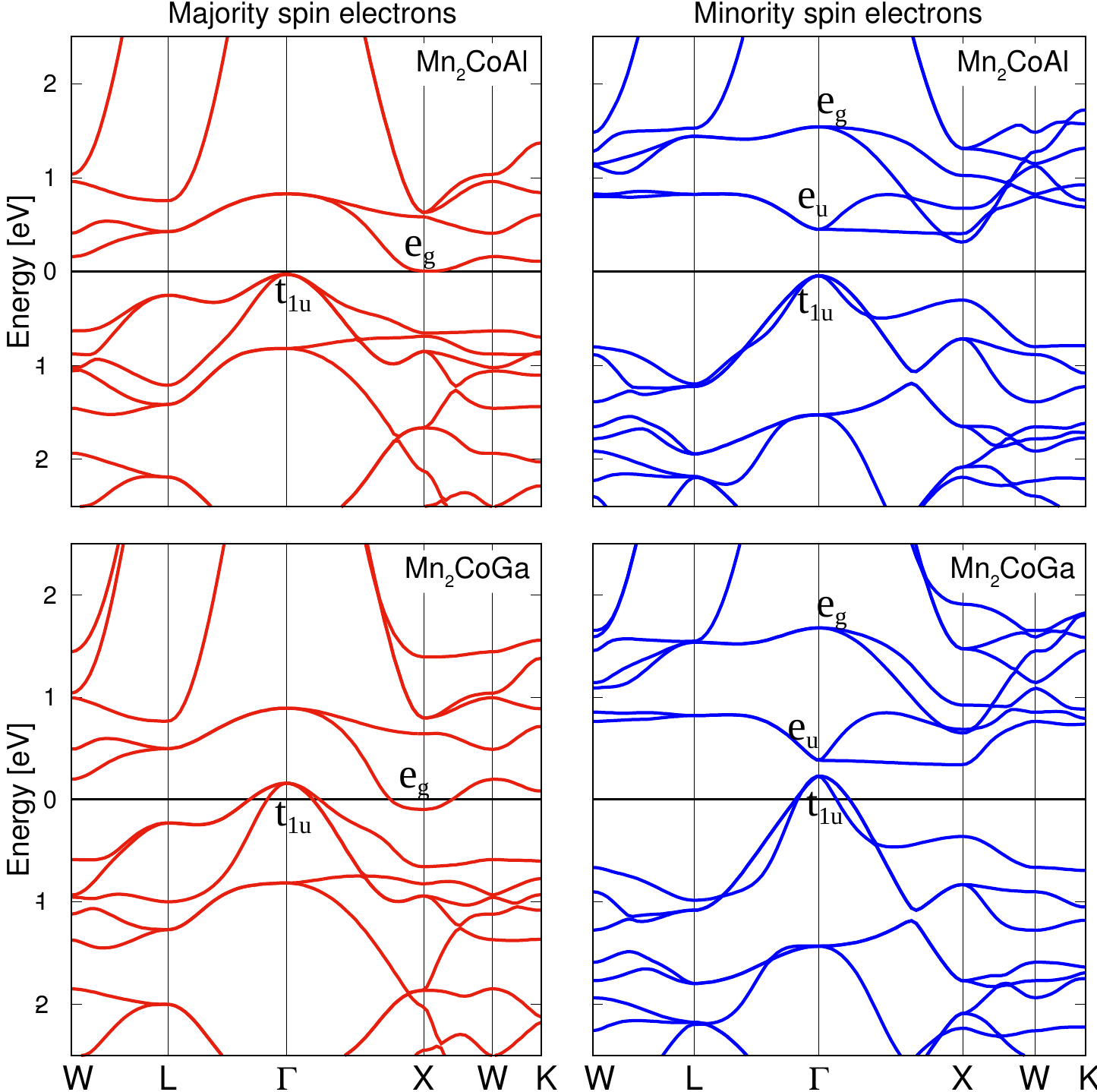}
	\caption{Calculated band structure of Mn$_2$CoAl and Mn$_2$CoGa for Majority(red) and Minority(blue) channel at lattice constant of  5.6506\AA \,  and  5.6695\AA \,   respectively. }
	
\end{figure}

 Fig.~\ref{Fig:1} represents the spin resolved band structure of Mn$_2$CoZ compounds near the Fermi level. The meticulous explanation of band structure can be done by invoking the hybridization scheme \cite{15} suggested by Galanakis \textit{et al.} for inverse Heusler alloys. The knowledge of Slater Pauling rule of full Heusler structure \cite{16} would be of great assistance while we are dealing with the hybridization scheme of inverse structure.
\par The close look of spin up channel of Mn$_2$CoAl around the fermi level shows that t$_{1u}$ and e$_g$ bands just touch the fermi level at point $\mathit{\Gamma}$ and \textit{X} respectively making it a strong candidate for Spin Gapless material. The spin down channel shows usual indirect band gap around the Fermi level. The gap of 0.358 eV is measured between the upper valence band at $\mathit{\Gamma}$  and lower conduction band at \textit{X}.        

{

\begin{table}[]

	\caption{Formation energy and cohesive energy of compounds under study.}
	\label{tab:tableis}
	\begin{tabular}{|l|c|c|}
		\hline
		Alloy                         & E$_{form}$(eV/atom) & E$_{coh}$(eV/atom) \\ \hline\hline
		Mn$_2$CoAl                    & -0.286              & -5.000             \\ \hline
		Mn$_2$CoCr$_{0.25}$Al$_{0.5}$ & -0.190              & -4.956             \\ \hline
		Mn$_2$CoCr$_{0.5}$Al$_{0.5}$  & -0.123              & -4.941             \\ \hline
		Mn$_2$CoFe$_{0.25}$Al$_{0.5}$ & -0.206              & -5.038             \\ \hline
		Mn$_2$CoFe$_{0.5}$Al$_{0.5}$  & -0.132              & -5.082             \\ \hline
		Mn$_2$CoGa                    & -0.192              & -4.725             \\ \hline
		Mn$_2$CoCr$_{0.25}$Ga$_{0.5}$ & -0.123              & -4.754             \\ \hline
		Mn$_2$CoCr$_{0.5}$Ga$_{0.5}$  & -0.078              & -4.806             \\ \hline
		Mn$_2$CoFe$_{0.25}$Ga$_{0.5}$ & -0.133              & -4.829             \\ \hline
		Mn$_2$CoFe$_{0.5}$Ga$_{0.5}$  & -0.079              & -4.938             \\ \hline
	\end{tabular}
\end{table}
}
In case of Mn$_2$CoGa, the metallic nature is dominant for majority spin electrons (see Fig.~\ref{Fig:1}) due to the intersection of t$_{1u}$ and e$_g$ bands around the Fermi level  while the band gap of 0.113 eV is measured for spin down channel. The valence band maximum i.e. t$_{1u}$ lies slightly above the Fermi level indicating weak half metallic character of Mn$_2$CoGa. In both cases, the band structure in the energy range of -2.5eV to 0eV are mainly due to the \textit{d}-electrons of the Manganese and Cobalt atoms. The scattered nature of bands in this region is predominantly due to the the strong hybridization between Manganese-Cobalt and Manganese-Manganese \textit{d}-electrons but there is also the contribution of \textit{p}-electrons from Al and Ga atoms. The low-lying s bands are not shown in the figure.
\par Since, by the transformation rule \cite{15}\cite{16}, t$_{1u}$ and e$_u$ states do not contain any \textit{d}-orbitals of Mn(B) atoms, the d hybrids of these states  are localized on Mn(A) and Co atoms. This is the reason why the states around the Fermi level are localized on Mn(A) and Co atoms. The gap between Mn(A) and Co is d-d band gap due to e$_u$-t$_{1u}$ splitting.
\par
\begin{figure}[h] 
	\centering
	\label{Fig:2}
	\includegraphics[width=1\linewidth, height=0.3\textheight]{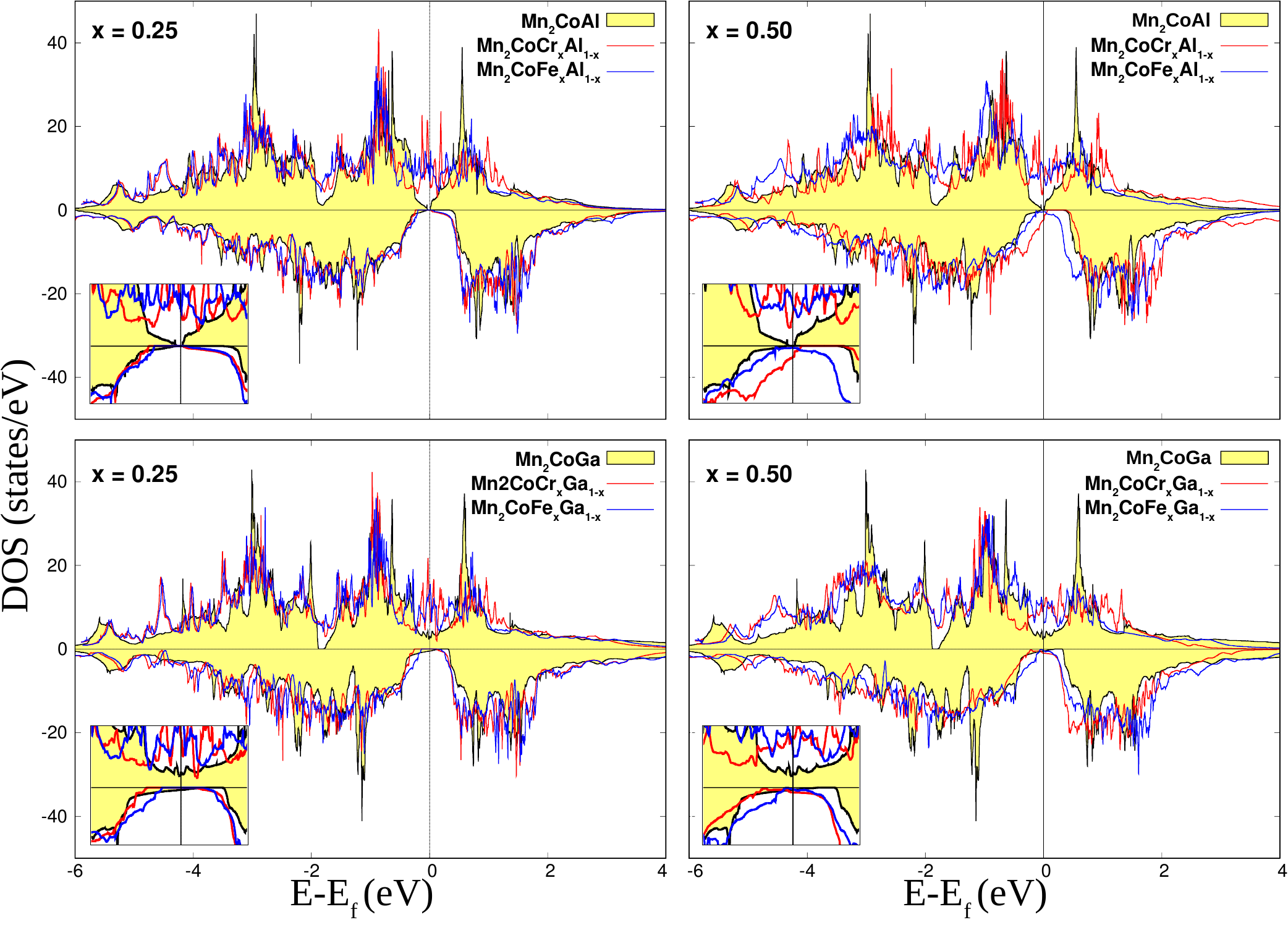}	
	\caption{Spin resolved total density of states(DOS) of Mn$_2$CoAl and Mn$_2$CoGa  for conventional
		cell at 5.6506\AA \, and 5.6695\AA \,  respectively. Blue line represents Fe doping while red line denotes Cr doping on Mn$_2$CoZ (Z=Al, Ga) compounds.}
\end{figure}

 From Fig.~\ref{Fig:2}, in case of pristine Mn$_2$CoAl, we can see that the dos around the Fermi level shows valley approaching zero peak indicating zero width energy gap supporting the SGS nature of band structure. For minority charge carriers,
there are no states around the fermi level, suggesting its semi conducting behavior. The dos of Mn$_2$CoGa shows the metallic nature of majority electrons whereas very small amount of density of states appeared around the Fermi level for the minority states. This indicates that the pristine Mn$_2$CoGa  may not have 100\% spin polarization at Fermi level, as in the case of half-metal.
\begin{table*}
	\caption{\label{tab:table1} Individual and total spin magnetic moments of Mn$_2$CoAl compound in case of Fe and Cr doping in $\mu_B$. * represents the changed value of spin magnetic moments of individual atoms after doping.}
		\begin{tabular}{|c|c|c|c|c|c|c|c|c|c|c|} 
			\hline
			Compounds & Mn(A)& Mn(B)& Mn*(B)& Co(C)& Co*(C)& Cr(A)& Fe(A)& Mn(D)& Al(D)&Total\\
			\hline
			Mn$_2$CoAl & -1.1996 & 2.3152 & - & 0.9379 & - & - & - & - & -0.03109& 8.00 \\ 
			\hline
			Mn$_2$CoCr$_{0.25}$Al$_{0.75}$ & -0.9640 & 2.6897 & 2.1677 & 1.1214 & 1.0975 & -1.3381 & - & 1.8923 & -0.0356& 11.00   \\ 
			\hline
			Mn$_2$CoCr$_{0.5}$Al$_{0.5}$ & -0.8526 & 2.5303 & 1.9578 & 1.2969 & 1.2810 & -1.2959 & - & 1.9786 & -0.0417& 13.56   \\ 
			\hline
			Mn$_2$CoFe$_{0.25}$Al$_{0.75}$ & -1.1586 & 2.7074 & 2.1376 & 1.1532 & 0.9793 & - & 1.0844 & 2.0216 & -0.0388& 13.00 \\ 
			\hline
			Mn$_2$CoFe$_{0.5}$Al$_{0.5}$ & -1.4142 & 2.4943 & 2.0041 & 1.3679 & 1.1740 & - & 1.2164 & 2.1179 & -0.0513&17.82  \\
			\hline
		\end{tabular}
\end{table*}
\begin{table*}
	\caption{\label{tab:table2}Individual and total spin magnetic moments of Mn$_2$CoGa compound in case of Fe and Cr doping in $\mu_B$. * represents the changed value of spin magnetic moments of individual atoms after doping.}
		\begin{tabular}{|c|c|c|c|c|c|c|c|c|c|c|} 
			\hline
			Compounds & Mn(A)& Mn(B)& Mn*(B)& Co(C)& Co*(C)& Cr(A)& Fe(A)& Mn(D)& Ga(D)&Total\\
			\hline
			Mn$_2$CoGa & -1.3611 & 2.4897 & - & 0.9131 & - & - & - & - & -0.0152&8.14   \\ 
			\hline
			Mn$_2$CoCr$_{0.25}$Ga$_{0.75}$ & -1.2148 & 2.8293 & 2.3434 & 1.1164 & 1.0592 & -1.4877 & - & 2.1053 & -0.0204&11.00   \\ 
			\hline
			Mn$_2$CoCr$_{0.5}$Ga$_{0.5}$ & -1.0929 & 2.6714 & 2.1038 & 1.3117 & 1.2535 & -1.4223 & - & 2.1292 & -0.0257&13.79   \\ 
			\hline
			Mn$_2$CoFe$_{0.25}$Ga$_{0.75}$ & -1.4142 & 2.8055 & 2.2731 & 1.1515 & 0.8991 & - & 1.1740 & 2.1408 & -0.0254&13.00   \\ 
			\hline
			Mn$_2$CoFe$_{0.5}$Ga$_{0.5}$ & -1.5800 & 2.6573 & 2.0072 & 1.3700 & 1.1351 & - & 1.1975 & 2.1959 & -0.0404&17.95   \\ 
			\hline
		\end{tabular}
\end{table*}
Now, let us focus on the band gap of Mn$_2$CoZ (Z=Al, Ga) compounds. We can see clearly that the band gap of pristine Mn$_2$CoAl is larger than the gap width of Mn$_2$CoGa. Liu \textit{et al.}\cite{2} have shown that smaller the value of lattice constant larger will be the band gap and vice versa, which is in good agreement with our case. The gap width largely depends upon the hybridization of delocalized \textit{p}-electrons from the Z atoms and localized \textit{d}-electrons of Manganese and Cobalt atoms. Since, the binding energy of \textit{p} electrons of Al is more than that of Ga atoms, the gap width of Mn$_2$CoAl is larger than that of Mn$_2$CoGa.
\par The blue line indicates Fe doping while the red line represents Cr doping in  Fig.~\ref{Fig:2}. The changes around the Fermi level is interesting and the modification in band gap is important for the study of different phenomena. For 25\% doping, the careful inspection around the Fermi level clearly shows that the SGS property of Mn$_2$CoAl has been destroyed and  weak-half metallic nature of  Mn$_2$CoGa has disappeared introducing half-metallic nature in both cases making doped  Mn$_2$CoZ compound a strong candidate of half-metal.
\par For Mn$_2$CoAl, 25\% doping of Fe or Cr decrease the band gap in comparison to its pristine form shifting the band gap on the left side but somehow half-metallicity is preserved. The band gap is measured to be 0.243 eV  and 0.254eV for Fe and Cr doping respectively which is smaller than the minority band gap in pure Mn$_2$CoAl. The narrowing of band gap is more in case of Fe doping than Cr doping. This can be explained by invoking the exchange splitting and coulomb repulsion competition, as we go on adding the electrons in the system through Fe doping. In case of Fe doping, in order to preserve half metallicity, the extra electron from Fe must occupy
anti-bonding majority states. Since these states have relatively higher energy than minority states, energetically it is not very approbative. Thus, the new states appears on the unoccupied minority states, the state which we can see in Fig.~\ref{Fig:2} near the edges of the gap, reducing the gap width of doped Mn$_2$CoAl.
\par In case of Mn$_2$CoGa, the doping gives different result. For 25\% doping of Fe or Cr, increase in the gap width is noticed unlike Mn$_2$CoAl. The gap width is 0.307eV for Fe doping while the value increased to 0.352eV for Cr doping. This peculiar behavior of Mn$_2$CoGa is beyond the scope of this paper and is the matter of further investigation. We hope this anomalous behavior will attract the attention of other researcher in this field.
\par In table \ref{tab:table1} and \ref{tab:table2} we have presented the total and individual spin magnetic moments of the Mn$_2$CoZ compounds. From table, it is clear that the ground state magnetic structure of Mn$_2$CoZ compounds is ferri-magnetic. Magnetic moment of Mn and Co at B and C sites are anti-ferromagnetically aligned with that of Mn and Z at A and D sites.  It is important to note that the Manganese atom at B site has highest value of spin magnetic moment which ferromagnetically couples with Cobalt atoms whose spin moment is near to unity. The reason for the ferromagnetic and antiferromagnetic alignment of different atoms in inverse Heusler alloy is described in reference \cite{2}.
\par In case of Mn$_2$CoAl, the total magnetic moment follows the Slater-Pauling rule, the rule which is important to analyze the electronic properties by studying the magnetic properties. The total number of valence electrons in Mn$_2$CoAl is 26, which is the sum of spin up and spin down electrons. Among 26 electrons, the 12 are in spin down states whereas 14 are in spin up states. These two extra electrons in spin up states are responsible for the magnetic moment of Mn$_2$CoAl. The conventional cell in this case is four times the primitive cell, so there would be total 8 extra electrons in spin up states giving total spin magnetic moment of 8$\mu_B$. But for Mn$_2$CoGa, as the upper valence state of minority electrons just cross the Fermi level (see Fig.~\ref{Fig:1}), we do not get the exact integer value for total spin magnetic moment, which can be verified from the table.
\par It is interesting to note that for 25\% doping, in all cases, there is an integer value of total spin magnetic moment, the strong evidence of the inducement of half-metallic properties in the doped samples. The non-integer value of total spin magnetic moment of 50\% doped samples support the fact that half-metallic properties is destroyed for large degree of doping.
\par From table \ref{tab:table1} and \ref{tab:table2}, we can see that the  spin magnetic moment alignment of  Cr(A) and Mn(A) is in same direction whereas in case of Fe doping unlike Cr(A), Fe(A) and Mn(A) are anti-feromagnetically aligned. This is the reason why the total spin magnetic moment value in Fe doped samples is more than for the Cr doped samples. Here, it has to be noted that in every doped sample the value of total spin magnetic moments is more than that of pristine compound. The observed increase in magnetic moment value in doped sample could be explained by taking into account the change in local environment of certain individual atoms by the
disorder produced from dopant atoms. The change in the magnetic moment of particular individual atoms sum up ultimately giving the increased value of magnetic moment of doped sample. Particularly, when we dope Fe or Cr, it occupies A site in accordance with general occupation rule and displace the Mn(A) atoms. These displaced Mn(A) atoms occupy the D sites and while doing so the anti-ferromagnetic alignment of Mn(A) atoms change to ferromagnetic alignment which largely increase the total spin magnetic moment value. Thus, the lion\textquoteright s share of credit for the increase in total magnetic moment goes to the Manganese atom. 

\section{Conclusion}
In our work, we have carried out the first principle study to investigate the effect of doping on SGS and weak half metallic properties of inverse Heusler alloys. For this, we have taken Mn$_2$CoZ (Al,Ga) compounds where Fe and Cr atoms were doped at two different concentration. We have found that the low degree doping enhance the half metallic properties in inverse Heusler alloys where as doping beyond certain degree destroys half-metallicity. It has been found that the Fe doping decreases the gap width where as Cr doping increase the band gap.The Magnetic moments are found to be drastically increases on doping concentrations accordingly.  
	     
\bibliography{article_citations}
\end{document}